\documentclass[twocolumn,english,aps,prl,showpacs]{revtex4}
\usepackage[T1]{fontenc}
\usepackage[latin1]{inputenc}
\usepackage{amsmath}
\usepackage{graphicx}
\usepackage{amssymb}
\usepackage{esint}

\makeatletter
\@ifundefined{textcolor}{}
{%
 \definecolor{BLACK}{gray}{0}
 \definecolor{WHITE}{gray}{1}
 \definecolor{RED}{rgb}{1,0,0}
 \definecolor{GREEN}{rgb}{0,1,0}
 \definecolor{BLUE}{rgb}{0,0,1}
 \definecolor{CYAN}{cmyk}{1,0,0,0}
 \definecolor{MAGENTA}{cmyk}{0,1,0,0}
 \definecolor{YELLOW}{cmyk}{0,0,1,0}
 }

\@ifundefined{definecolor}
 {\usepackage{color}}{}
\makeatother

\makeatother

\usepackage{babel}

\makeatother

\usepackage{babel}

\makeatother

\usepackage{babel}

\begin{document}

\title{Quantitative comparison between theoretical predictions and experimental
results for Bragg spectroscopy of a strongly interacting Fermi superfluid }

\author{Peng Zou$^{1,2}$, Eva D. Kuhnle$^{2}$, Chris J. Vale$^{2}$, and
Hui Hu$^{2}$}

\affiliation{$^{1}$Department of Physics, Renmin University of China, Beijing
100872, China}

\affiliation{$^{2}$ARC Centre of Excellence for Quantum-Atom Optics and Centre
for Atom Optics and Ultrafast Spectroscopy, Swinburne University of
Technology, Melbourne 3122, Australia}

\date{\today{}}
\begin{abstract}
Theoretical predictions for the dynamic structure factor of a harmonically
trapped Fermi superfluid near the BEC-BCS crossover are compared with
recent Bragg spectroscopy measurements at large transferred momenta.
The calculations are based on a random-phase (or time-dependent Hartree-Fock-Gorkov)
approximation generalized to the strongly interacting regime. Excellent
agreement with experimental spectra at low temperatures is obtained,
with no free parameters. Theoretical predictions for zero-temperature
static structure factor are also found to agree well with the experimental
results and independent theoretical calculations based on the exact
Tan relations. The temperature dependence of the structure factors
at unitarity is predicted. 
\end{abstract}

\pacs{05.30.Fk, 03.75.Hh, 03.75.Ss, 67.85.-d}

\maketitle
Ultracold Fermi gases of $^{6}$Li and $^{40}$K atoms near Feshbach
resonances provide a new paradigm for studying strongly correlated
many-body systems \cite{bloch}. At low temperatures, they display
the intriguing crossover from a Bose-Einstein condensate (BEC) to
a Bardeen-Cooper-Schrieffer (BCS) superfluid \cite{giorgini}. In
the unitarity regime at the cusp of the crossover, a superfluid with
neither dominant bosonic nor fermionic character emerges that exhibits
universal properties that might be found in other strongly interacting
superfluids \cite{ho,hdlNatPhys}, such as high-temperature superconductors
or nuclear matter in neutron stars. This new superfluid has already
been investigated intensively \cite{bloch,giorgini}, leading to several
milestone observations, some of which still defy theoretical understanding.
Here we present a quantitative description of the recent two-photon
Bragg spectroscopy measurement for this new superfluid \cite{sutbragg}.

Theoretical challenges in describing the BEC-BCS crossover arise from
its strongly correlated nature: there is no small interaction parameter
to set the accuracy of theories \cite{hldNJP}. Significant progress
has been made in developing better quantum Monte Carlo simulations
\cite{astrakharchik,bulgac,burovski,carlson} and strong-coupling
theories \cite{hldNJP,ohashi,hldepl,lhpra,haussmann}, leading to
the quantitative establishment of a number of properties. These include
equation of state \cite{astrakharchik,hldNJP,hldepl,dukeEoS,ensEoS,ensEoS2},
frequency of collective oscillations \cite{huprl2004,altmeyer}, pairing
gap \cite{carlson,schunck}, and superfluid transition temperature
\cite{burovski,horikoshi}. However, other fundamental properties,
such as the single-particle spectral function measured by rf spectroscopy
\cite{gaebler,huprl2010} and the dynamic structure factor probed
by Bragg spectroscopy \cite{sutbragg}, are not as well understood.

\begin{figure}[htp]

\begin{centering}
\includegraphics[clip,width=0.48\textwidth]{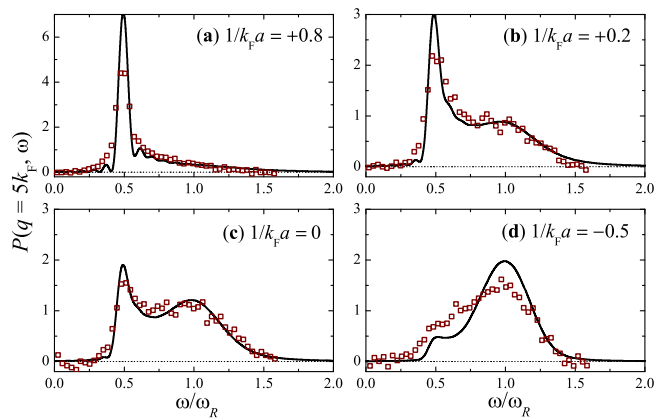} 
\par\end{centering}

\caption{(color online). Quantitative comparison of theoretical and experimental
Bragg spectra {[}see Eq. (\ref{Pimp}){]}. The RPA prediction (lines)
agrees well with the experimental data (empty squares) \cite{sutbragg}
at the BEC-BCS crossover, with no free parameters. The spectrum is
normalized so that the area below the curve is unity. The frequency
is measured in units of the recoil energy of the atoms (see text).}

\label{fig1} 
\end{figure}

In this Rapid Communication, we show that a random-phase approximation
(RPA), generalized to the strongly interacting regime, is able to
describe quantitatively the observed Bragg spectra for harmonically
trapped $^{6}$Li atoms at large transferred momenta. This surprising
result indicates that the RPA captures the essential physics and constitutes
a reasonable approximation for the strongly interacting region of
the BEC-BCS crossover, particularly the low temperature range accessed
by most experiments. The RPA method has previously been used to study
the dynamic structure factor \cite{minguzzi} and collective oscillations
\cite{bruun} of weakly interacting Fermi superfluids. A dynamic mean-field
approach, identical to the RPA but based on kinetic equations, was
developed to investigate structure factors \cite{combescot} and collective
modes \cite{combescotPRA} of a uniform, strongly interacting Fermi
gas. At finite temperatures, structure factors at the crossover were
also studied using a pseudogap theory \cite{guo}.

Our main result is summarized in Fig. 1, which shows the normalized
experimental Bragg spectra \cite{sutbragg} along with the RPA predictions.
Excellent agreement is found, with no free parameters.

We begin by outlining briefly the RPA using the Hamiltonian (hereafter
$\hbar=1$),

\begin{eqnarray}
{\cal H} & = & \sum_{\sigma}\int d{\bf r}\psi_{\sigma}^{+}({\bf r})\left[-\frac{\nabla^{2}}{2M}-\mu+V_{T}({\bf r})\right]\psi_{\sigma}({\bf r})\nonumber \\
 &  & +U_{0}\int d{\bf r}\psi_{\uparrow}^{+}({\bf r})\psi_{\downarrow}^{+}({\bf r})\psi_{\downarrow}({\bf r})\psi_{\uparrow}({\bf r}),\label{hami}\end{eqnarray}
 which describes a balanced spin-1/2 ($\sigma=\uparrow,\downarrow$)
Fermi gas with mass $M$ in a harmonic trap $V_{T}({\bf r})$, where
fermions with unlike spins interact via a contact potential $U_{0}\delta({\bf r-r}^{\prime})$.
The total number of atoms $N$ is tuned by the chemical potential
$\mu$ and the bare interaction strength $U_{0}$ is renormalized
by the \textit{s}-wave scattering length $a$, $1/U_{0}+\sum_{{\bf k}}M/{\bf k}^{2}=M/(4\pi a)$.
In the superfluid phase, we treat the system as a gas of long-lived
Bogoliubov quasiparticles interacting through a mean-field and consider
its response to a weak external field of the form of $\delta Ve^{i({\bf q\cdot r}-\omega t)}$.
The essential idea of the RPA is that there is a self-generated mean-field
potential experienced by quasiparticles \cite{liupra2004}, associated
with the local changes in the density distribution of the two spin
species, $\delta U=U_{0}\int d{\bf r(}\sum_{\sigma}\delta n_{\sigma}\psi_{\sigma}^{+}\psi_{\sigma}+\delta m\psi_{\uparrow}^{+}\psi_{\downarrow}^{+}+\delta m^{*}\psi_{\downarrow}\psi_{\uparrow})$,
where $\delta n_{\sigma}\equiv\delta n_{\sigma}\left({\bf r},t\right)$
and $\delta m\equiv\delta m\left({\bf r},t\right)$ are the normal
and anomalous density fluctuations, respectively, which must be determined
self-consistently. In the linear approximation, the self-generated
potential $\delta U$ plays the same role as the perturbation field
when we calculate the dynamic response using a static BCS Hamiltonian
as the reference system \cite{minguzzi,bruun,liupra2004}. This leads
to coupled equations for density fluctuations. The linear response
is characterized by a matrix consisting of all two-particle response
functions: \[
\chi\equiv\left\{ \begin{array}{cccc}
\langle\langle\hat{n}_{\uparrow}\hat{n}_{\uparrow}\rangle\rangle & \langle\langle\hat{n}_{\uparrow}\hat{n}_{\downarrow}\rangle\rangle & \langle\langle\hat{n}_{\uparrow}\hat{m}\rangle\rangle & \langle\langle\hat{n}_{\uparrow}\hat{m}^{+}\rangle\rangle\\
\langle\langle\hat{n}_{\downarrow}\hat{n}_{\uparrow}\rangle\rangle & \langle\langle\hat{n}_{\downarrow}\hat{n}_{\downarrow}\rangle\rangle & \langle\langle\hat{n}_{\downarrow}\hat{m}\rangle\rangle & \langle\langle\hat{n}_{\downarrow}\hat{m}^{+}\rangle\rangle\\
\langle\langle\hat{m}\hat{n}_{\uparrow}\rangle\rangle & \langle\langle\hat{m}\hat{n}_{\downarrow}\rangle\rangle & \langle\langle\hat{m}\hat{m}\rangle\rangle & \langle\langle\hat{m}\hat{m}^{+}\rangle\rangle\\
\langle\langle\hat{m}^{+}\hat{n}_{\uparrow}\rangle\rangle & \langle\langle\hat{m}^{+}\hat{n}_{\downarrow}\rangle\rangle & \langle\langle\hat{m}^{+}\hat{m}\rangle\rangle & \langle\langle\hat{m}^{+}\hat{m}^{+}\rangle\rangle\end{array}\right\} ,\]
 where $\langle\langle\hat{A}\hat{B}\rangle\rangle$ is the Fourier
transform of the retarded function $-i\Theta\left(t-t^{\prime}\right)\langle[\hat{A}({\bf r},t),\hat{B}({\bf r}^{\prime},t^{\prime})]\rangle$.
For simplicity, we abbreviate $\chi_{\sigma\sigma^{\prime}}\equiv\langle\langle\hat{n}_{\sigma}\hat{n}_{\sigma^{\prime}}\rangle\rangle$,
$\chi_{\sigma m}\equiv\langle\langle\hat{n}_{\sigma}\hat{m}\rangle\rangle$,
$\chi_{\sigma\bar{m}}\equiv\langle\langle\hat{n}_{\sigma}\hat{m}^{+}\rangle\rangle$,
$\chi_{m\bar{m}}\equiv\langle\langle\hat{m}\hat{m}^{+}\rangle\rangle$,
and so on. By solving the coupled equations for density fluctuations,
the standard RPA response function $\chi$ can be expressed in terms
of the static BCS response function $\chi^{0}$ \cite{bruun}, \begin{equation}
\chi=\chi^{0}\left[\hat{1}-U_{0}\chi^{0}{\cal G}\right]^{-1},\label{RPA}\end{equation}
 where ${\cal G}=\delta({\bf r}-{\bf r}^{\prime})[\sigma_{0}\otimes\sigma_{x}]$
is a direct product of two Pauli matrices $\sigma_{0}$ and $\sigma_{x}$
and the unit matrix $\hat{1}=\delta({\bf r}-{\bf r}^{\prime})[\sigma_{0}\otimes\sigma_{0}]$.
The dynamic structure factor $S_{\sigma\sigma^{\prime}}(\omega)$
is related to the normal density response function by the fluctuation-dissipation
theorem, \begin{equation}
S_{\sigma\sigma^{\prime}}(\omega)=-\frac{1}{\pi}\frac{1}{\left[1-\exp\left(-\omega/k_{B}T\right)\right]}\mathop{\rm Im}\chi_{\sigma\sigma^{\prime}}\left(\omega\right),\end{equation}
 and the static structure factor is given by $S_{\sigma\sigma^{\prime}}=(2/N)\int_{-\infty}^{+\infty}d\omega S_{\sigma\sigma^{\prime}}(\omega)$.
In the weak-coupling regime, Eq.~(\ref{RPA}) can be solved by calculating
$\chi^{0}$ for a thermal average of BCS quasiparticles \cite{minguzzi,bruun}.

Here, we extend the RPA to the strongly interacting regime with an
arbitrarily large scattering length $a$, by properly renormalizing
the bare interaction strength $U_{0}$ and the two response functions
$\chi_{m\bar{m}}^{0}$ and $\chi_{\bar{m}m}^{0}$, which was found
to be suitable at the BEC-BCS crossover \cite{haussmann,pieri}. The
ultraviolet divergence of these two functions \cite{bruun} is canceled
exactly by the small value of $U_{0}$, when the momentum cut-off
goes to infinity. In homogeneous systems, a careful account of the
divergent terms in the inverted matrix of the RPA equation (\ref{RPA})
leads to a concise expression for the response functions: \begin{eqnarray}
\chi_{\uparrow\uparrow} & = & \chi_{\uparrow\uparrow}^{0}-\left[2\chi_{\uparrow\downarrow}^{0}\chi_{\uparrow m}^{0}\chi_{\uparrow\bar{m}}^{0}+\left(\chi_{\uparrow m}^{0}\right)^{2}\tilde{\chi}_{m\bar{m}}^{0}\right.\nonumber \\
 &  & \left.+\left(\chi_{\uparrow\bar{m}}^{0}\right)^{2}\tilde{\chi}_{\bar{m}m}^{0}\right]/\left[\tilde{\chi}_{m\bar{m}}^{0}\tilde{\chi}_{\bar{m}m}^{0}-\left(\chi_{\uparrow\downarrow}^{0}\right)^{2}\right],\label{RPAupup}\end{eqnarray}
 and \begin{equation}
\chi_{\uparrow\downarrow}=\chi_{\uparrow\uparrow}-\chi_{\uparrow\uparrow}^{0}+\chi_{\uparrow\downarrow}^{0},\label{RPAupdw}\end{equation}
 where the response functions with a tilde, i.e., $\tilde{\chi}_{m\bar{m}}^{0}\equiv\chi_{m\bar{m}}^{0}+\sum_{{\bf k}}M/{\bf k}^{2}-M/(4\pi a)$,
become free from any ultraviolet divergence. The above equations were
previously obtained by Combescot and collabrators using kinetic equations
(see Eq. (B22) in Ref. \cite{combescotPRA}). Note that, we use a 
Leggett-BCS ground state without inclusion of the Hartree-Fock term in 
the quasiparticle spectrum. Therefore, in the BCS regime our treatment 
does not account for the leading interaction effect as in Refs. \cite{minguzzi,bruun}. 
At the crossover, however, it does capture the dominant pairing gap. 
Note also that, the RPA method accounts for single particle-hole excitations. 
Higher correlations such as multi-particle-hole excitations are neglected.

In the presence of a harmonic trap, the renormalization procedure
becomes cumbersome because of the discrete energy levels. It is convenient
to use a local density approximation (LDA) that treats the system
as a collection of many homogeneous cells with local chemical potential
\cite{giorgini}, $\mu({\bf r})=\mu-V_{T}({\bf r})$, where $V_{T}({\bf r})=M(\omega_{x}^{2}x^{2}+\omega_{y}^{2}y^{2}+\omega_{z}^{2}z^{2}{\bf )}/2$
is the harmonic trapping potential. The LDA treatment is valid for
a large number of atoms such as $N\sim10^{5}$ as in experiments.
It has been used extensively in studying the static density profile
of either atomic Fermi, Bose gases \cite{LDAFG} or Bose-Fermi mixtures
\cite{LDABF}. In the nuclear context, it has also been used to calculate
the dynamic response function \cite{LDASchuck}. At a given temperature
and scattering length, we solve the Leggett-BCS equation with local
chemical potential for the local pairing gap and calculate the static
response function $\chi^{0}$, then solve the local RPA density response
functions using Eqs. (\ref{RPAupup}) and (\ref{RPAupdw}), and finally
obtain the total RPA responses by integrating over the whole trap.
In our calculations, the interaction strength is characterized by
the dimensionless parameter, $1/(k_{F}a)$, where $k_{F}=\sqrt{2ME_{F}}$
is the Fermi wave vector and the Fermi energy is $E_{F}=(3N\omega_{x}\omega_{y}\omega_{z})^{1/3}$.

\begin{figure}[htp]
\begin{centering}
\includegraphics[clip,width=0.4\textwidth]{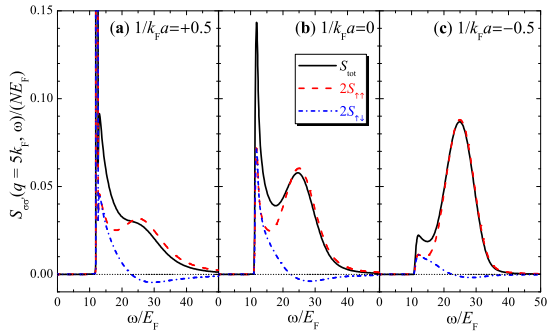} 
\par\end{centering}

\caption{(color online). Zero temperature spin parallel $S_{\uparrow\uparrow}({\bf q},\omega)$
(dashed lines), anti-parallel $S_{\uparrow\downarrow}({\bf q},\omega)$
(dot-dashed lines), and total dynamic structure factor $S({\bf q},\omega)=2[S_{\uparrow\uparrow}+S_{\uparrow\downarrow}]$
(solid lines) across the BEC-BCS crossover: $1/k_{F}a=0.5$ (a), $0.0$
(b), and $-0.5$ (c). The negative weight in $S_{\uparrow\downarrow}$
at about the recoil energy is consistent with the exact sum rule $\int\omega S_{\uparrow\downarrow}(\mathbf{q},\omega)d\omega=0$
\cite{guo}.}

\label{fig2} 
\end{figure}

Figure 2 shows the zero-temperature spin parallel, anti-parallel,
and total dynamic structure factor at a transferred wave-vector $q=5k_{F}$
in the BEC-BCS crossover, calculated using the above RPA procedure
for a trapped Fermi gas. In addition to a broad response at the recoil
energy $\omega_{R}=q^{2}/(2M)=25E_{F}$ caused by resonant scattering
of atoms, a much narrower peak develops at about $\omega_{R}/2$ with
increased coupling. The peak, commonly referred to as the quasielastic
peak in the literature, is found by the recent theoretical calculation
\cite{combescot} and the observation of Bragg spectroscopy \cite{sutbragg}.
This is simply the Bogoliubov-Anderson phonon mode of a Fermi superfluid
at large wave-vectors, which evolves continuously into a Bogoliubov
mode of molecules towards the BEC limit \cite{combescot}. The molecular
peak is mostly evident in $S_{\uparrow\downarrow}$ as there is no
background atomic response. Measurments of $S_{\uparrow\downarrow}$
may also help establish the presence of Fermi superfluidity \cite{guo}.

To make a quantitative comparison with the experimental spectra, we
calculate the momentum imparted to the Fermi cloud, the quantity measured
directly in the Bragg scattering experiment \cite{sutbragg,brunello}:
\begin{equation}
{\cal P}\left({\bf q},\omega\right)\propto\frac{1}{\pi\sigma}\int_{-\infty}^{\infty}d\omega^{\prime}S({\bf q},\omega^{\prime})\mathrm{sinc^{2}}\left[\frac{\omega-\omega^{\prime}}{\sigma}\right],\label{Pimp}\end{equation}
 where $\mathrm{sinc}(x)=\sin(x)/x$ and the energy resolution $\sigma=2/\tau_{Br}$
is set by the experimental Bragg pulse duration ($\tau_{Br}=40$ $\mu$s)
\cite{sutbragg}. We find $\sigma\approx0.68E_{F}$ $\approx0.027\omega_{R}$.
Figure 1 presents a comparison of the experimental data (open squares)
with the RPA predictions (lines) for the Bragg spectra normalized
in such a way that $\int{\cal P}({\bf q},\omega)d\omega=1$. With
no free parameters, our RPA predictions agree well with the experimental
results in the unitarity regime ($1/k_{F}a=0.0$ and $0.2$) and BEC
regime ($1/k_{F}a=0.8$). The agreement on the BCS side ($1/k_{F}a=-0.5$),
however, becomes worse. The quantitative agreement around unitarity
is very compelling, since the RPA was assumed to be unreliable in
the (strongly interacting) regime of large pair fluctuations. Our
comparison indicates that the RPA is able to describe the dynamical
properties of the BEC-BCS crossover, at least at zero temperature
and large momenta. High order multi-particle-hole excitations, absent
in the RPA theory, seems to be negligibly small at large momenta.
More studies are needed to understand this. Note finally that, the
somewhat poorer agreement at $1/k_{F}a=-0.5$ can be attributed to
the mean-field shift, which is ignored in the RPA but dominates for
sufficiently weak interactions.

\begin{figure}[htp]
\begin{centering}
\includegraphics[clip,width=0.4\textwidth]{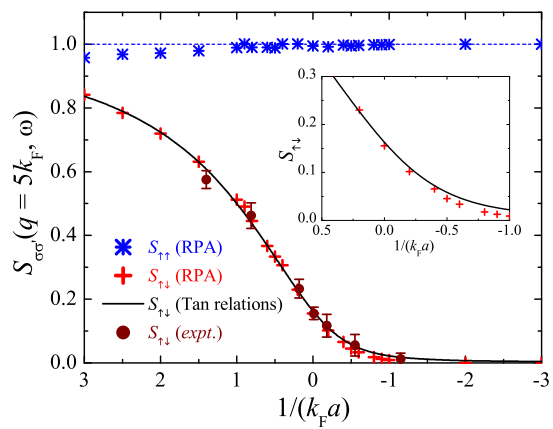} 
\par\end{centering}

\caption{(color online). Quantitative comparison between theory and experiment
for the zero temperature static structure factor across the crossover.
For $S{}_{\uparrow\downarrow}$, with no free parameters our RPA prediction
(plus symbols) agrees well with the experimental data for $S(\mathbf{q},\omega)-1$
(solid circles with error bars) \cite{sutTanRelation} and an independent
theoretical result based on the exact Tan relations (solid line) \cite{hldepl2010}.
At large transferred momentum, $S{}_{\uparrow\uparrow}\simeq1$. The
inset highlights the RPA prediction with respect to the Tan-relation
result in the BCS regime.}

\label{fig3} 
\end{figure}

The agreement between the RPA theory and the Bragg experiment is further
confirmed by comparing the spin anti-parallel static structure factor
at zero temperature, as reported in Fig. 3. Experimentally, the static
structure factor can be measured model-independently by invoking the
$f$-sum rule \cite{sutTanRelation}; while, theoretically, it can
be determined very accurately using the exact Tan relations and the
known equation of state \cite{hldepl2010}. It is evident from Fig.
3 that the RPA prediction fits very well with the experimental data,
as well as the independent theoretical result based on the Tan relations.
In particular, the two theoretical predictions are nearly indistinguishable
on the BEC side with $1/k_{F}a\geq0$. However, they differ towards
the BCS limit, as highlighted in the inset. The discrepancy is consistent
with Fig. 1d where the RPA predicts less pairing and hence lower $S(q)$.

\begin{figure}[htp]
\begin{centering}
\includegraphics[clip,width=0.4\textwidth]{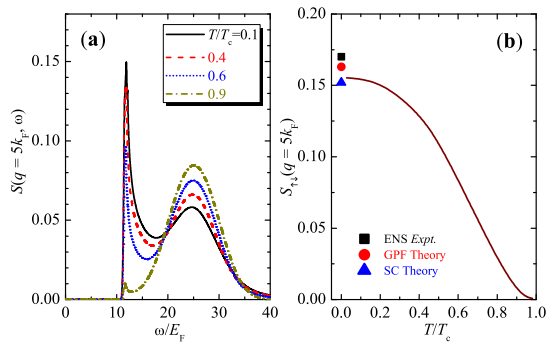} 
\par\end{centering}

\caption{(color online). Temperature dependence of the dynamic (a) and static
(b) structure factor for a unitary Fermi gas in harmonic traps at
$q=5k_{F}$. According to the Tan
relation, $S_{\uparrow\downarrow}({\bf q})\simeq128\zeta/[175\xi^{1/4}(q/k_{F})]$
at $T=0$ \cite{sutTanRelation}, where $\xi$ and $\zeta$ are the
universal parameters at unitarity \cite{ensEoS2}. The symbols in
(b) show predictions using theoretically or experimentally determined
$\xi$ and $\zeta$: ENS experiment (square) \cite{ensEoS2}, Gaussian
pair fluctuation theory (circle) \cite{hldepl}, and self-consistnent
theory (triangle) \cite{haussmann2}. Here, $T_{c}\simeq0.37T_{F}=0.37E_{F}/k_{B}$.}

\label{fig4} 
\end{figure}

A more stringent test of the RPA theory may be provided by the temperature
or momentum dependence of dynamic and static structure factors. In
Fig. 4, we predict the dynamic and static structure factor as a function
of temperature for a trapped Fermi gas at unitarity, which will be
investigated in future experiments. As anticipated, the pair (atomic)
response increases (decreases) with decreasing the temperature, leading
to a monotonic decay of the static structure factor.

The present RPA theory is most likely valid only in a narrow temperature
window near $T=0$. With increasing temperature, the pairing gap decreases
and thermal pair fluctuations increase. The RPA will eventually break
down at a characteristic temperature $T_{RPA}(\lesssim T_{c})$. This
is evident in Fig. 4b where the spin anti-parallel static structure
factor vanishes unphysically above the superfluid transition temperature.

At low transferred momenta, quantum fluctuations are likely to increase
and the RPA theory will become less reliable. To overcome these limitations,
we could use a Cooperon-mediated interaction (many-body $T$-matrix)
to replace the bare contact interaction \cite{buchler}, or use the
phenomenological Landau parameters for the mean-field shift \cite{stringari},
as determined from thermodynamic measurements \cite{ensEoS} or quantum
Monte Carlo simulations.

In summary, we have used a strong-coupling RPA theory to calculate
the dynamic and static structure factors of a trapped Fermi gas at
the BEC-BCS crossover. The theory is quantitatively applicable at
low temperatures and large transferred momenta, as confirmed by the
excellent agreement with the experimental Bragg spectra. The RPA theory
thus seems to provide a novel starting point for investigating dynamic
properties of a strongly interacting Fermi gas at finite temperatures
and low momenta.

We are indebted to P. Drummond, P. Hannaford, W. Zhang and N.-H. Tong
for fruitful discussions. Numerical calculations were performed at
the Physics Laboratory for High Performance Computing, Renmin University
of China. This work was supported by the ARC Centre of Excellence
for Quantum-Atom Optics, ARC Discovery Project Nos. DP0984522 and
DP0984637, and NSFC Grant No. 10774190. Correspondence should be addressed
to HH at hhu@swin.edu.au.

\end{document}